\documentclass[%
 reprint,
 amsmath,amssymb,
 aip,
 pop,
]{revtex4-2}

\usepackage{graphicx}
\usepackage{bm}
\usepackage{physics}
\usepackage{siunitx}
\usepackage{hyperref}
\usepackage{xcolor}
\usepackage{booktabs}
\usepackage{dcolumn}
\usepackage[utf8]{inputenc}
\usepackage[T1]{fontenc}
\usepackage{mathptmx}
\usepackage{etoolbox}

\makeatletter
\def\@email#1#2{%
 \endgroup
 \patchcmd{\titleblock@produce}
  {\frontmatter@RRAPformat}
  {\frontmatter@RRAPformat{\produce@RRAP{*#1\href{mailto:#2}{#2}}}\frontmatter@RRAPformat}
  {}{}
}%
\makeatother
\begin{document}

\title{Millisecond-Scale Neural Operator Surrogates for Double-Null Free-Boundary Grad--Shafranov Equilibria}

\author{Plamen G. Krastev}
\email{plamenkrastev@fas.harvard.edu}
\affiliation{Harvard University, Faculty of Arts and Sciences, Research Computing, 38 Oxford Street, Cambridge, MA 02138, USA}

\date{\today}

\begin{abstract}
The Grad-Shafranov (GS) equation is a key relation governing the ideal
magnetohydrodynamic (MHD) equilibrium in tokamak plasmas. One of the
major challenges in using free-boundary GS solvers in real-time control
workflows and large-scale optimization and design scenarios is the large
computational cost and sample-dependent latency of nonlinear Picard
iteration. In particular, repeated high-fidelity evaluations for diverted
tokamak equilibria can become prohibitive when the equilibrium computation
is inside an optimization, modeling, or control-oriented loop. Here we train
a geometrically conditioned Fourier Neural Operator (FNO) to learn a
computationally efficient representation of a constrained forward map from
spatial coordinates, scalar operating parameters
$(P_{\mathrm{axis}}, I_p, f_{\mathrm{vac}})$, and prescribed X-point
locations to the full poloidal-flux field $\psi(R,Z)$. The FNO model is
trained on a controlled family of constrained double-null free-boundary
equilibrium configurations generated with the \textsc{FreeGS} solver for a
single fixed machine geometry and prescribed topology. The best model
achieves a mean relative $L^2$ error of $0.05\%$, with the test error
following an empirical $N^{-0.68}$ power law over the range
$N_{\mathrm{train}}\in\{500,1000,2000,5000\}$, indicating systematic
improvement with training-set size over the tested regime. It also recovers
both X-points to within $0.2$~cm and localizes the O-point to
$0.03$~cm. As a physics-consistency diagnostic, the predicted fields satisfy
an external finite-difference GS residual evaluation at the same level as
the ground-truth fields, with a mean normalized residual of
$2.29$, indistinguishable from the $2.29\pm0.06$ baseline obtained on
\textsc{FreeGS} solutions using the same diagnostic. The trained FNO
surrogate evaluates a single equilibrium in $2.77$~ms on GPU and $25.6$~ms on
CPU, corresponding to speedups of ${\sim}640\times$ and
${\sim}69\times$ relative to \textsc{FreeGS} as configured in these
experiments, with near-deterministic latency (p95/median~$=1.01$). These
results demonstrate that neural-operator surrogates can provide accurate,
geometrically precise, and millisecond-scale equilibrium evaluations for
magnetic-confinement fusion workflows within a prescribed topology and
machine geometry.
\end{abstract}
\maketitle

\section{Introduction}

Plasma equilibrium in axisymmetric geometry in fusion devices is governed by the nonlinear elliptic Grad--Shafranov (GS) equation for the poloidal magnetic flux $\psi(R,Z)$ \cite{grad1958hydromagnetic,shafranov1958equilibrium,wesson2011tokamaks}. The resulting field determines the magnetic axis, flux-surface geometry, separatrix shape, and X-point structure, and accordingly it is critical for tasks such as stability analysis, transport modeling, equilibrium reconstruction, and plasma shape control \cite{freidberg2014idealMHD}. Accurate and rapid evaluation of the plasma equilibrium is therefore a central computational task in magnetic-confinement fusion.

A key challenge in free-boundary equilibrium modeling is that the plasma boundary is not known in advance. In diverted configurations, it must emerge self-consistently from the plasma profiles, external coil fields, and global geometric constraints. Therefore production codes compute nonlinear GS updates while enforcing constraints, including isoflux conditions and explicit X-point locations \cite{lao1985efit,ferron1998rtefit,moret2015liuqe}. These approaches are robust and mature, but they employ iterative nonlinear numerical algorithms with sample-dependent convergence latency, and since equilibria are evaluated many times, have non-negligible computational cost. Specifically, repeated high-fidelity GS evaluations can become prohibitive in many-query settings such as uncertainty quantification, outer-loop optimization, integrated modeling, and control-oriented workflows \cite{elman2024mlmc_gs}, while recent surrogate-equilibrium efforts such as EFIT-Prime (Equilibrium FIT) highlight the complementary need for fast, uncertainty-aware equilibrium reconstruction in real-time and control-relevant settings \cite{madireddy2024efitprime}. Real-time equilibrium reconstruction has been achieved in experimental settings through machine-specific optimization \cite{ferron1998rtefit,moret2015liuqe}, but such systems are typically engineered around the diagnostics, actuators, and temporal coherence of a specific device. On the other hand, emerging use cases such as digital twins, controller training, and rapid parametric scans require repeated evaluation across families of operating points for which warm-started iteration is either unavailable or impractical.

Machine learning (ML) has become an important tool in nuclear fusion research, with applications ranging from transport modeling and integrated simulation to disruption prediction and plasma control \cite{kates2021surrogate,meneghini2017selfconsistent,degrave2022magnetic,seo2021feedforward,2024NatCo..15.3990K,Wang2025RampDownTokamaks,2025NatCo..16.8877W,Wu2025HL3TokamakRL,Montes2019DisruptionWarnings,Rea2019RealTimeDisruptionPredictor,Rea2020InterpretableDisruptionPredictors,Zheng2023TransferLearningTokamaks,Tang2023AIDeepLearningDisruptionPredictor,Spangher2025DisruptionBench,Seo2024PhysicsInformedNeuralTransport,Gopakumar2024PlasmaSurrogateFNO,Carey2025NeuralOperatorEdgeSimulations,joung2020deep,Joung2023GSDeepNet,lu2023east,wai2022neural,mcclenaghan2024augmenting}. In particular, ML approaches have been applied to GS equilibria in several recent works. Physics-informed neural networks (PINNs) \cite{raissi2019pinn} have been used to solve the GS equation \cite{jang2024gs_pinn}, but a trained PINN model typically evaluates a single problem instance or a narrow family rather than providing solutions across a wide parametric set. Neural-network and reduced-order approaches have also been developed for fast full-field equilibrium reconstruction from magnetic diagnostics. Joung et al. developed a diagnostic-constrained neural-network Grad--Shafranov solver for KSTAR \cite{Park_2011} that maps measured magnetic signals and spatial coordinates to $\psi(R,Z)$ and $\Delta^\star\psi$, reproducing off-line EFIT-quality equilibria with approximately millisecond latency~\cite{joung2020deep}. Lu et al. similarly developed a deep-learning reconstruction model for EAST \cite{Yuanxi_2006} that maps magnetic measurements, coil currents, and plasma current to the two-dimensional poloidal flux $\Psi(R,Z)$ trained on off-line EFIT reconstructions~\cite{lu2023east}. GS-DeepNet further introduced a self-teaching physics-constrained framework for KSTAR equilibrium reconstruction based on neural representations of the Grad--Shafranov equation~\cite{Joung2023GSDeepNet}. Wai et al. developed Eqnet for NSTX-U \cite{Munaretto_2026}, a neural-network equilibrium model trained on EFIT01 reconstructions that can operate either in a diagnostic-reconstruction mode or in a forward mode using coil currents, vessel currents, and pressure and current-profile information to predict equilibrium flux surfaces~\cite{wai2022neural}. More recently, McClenaghan et al. augmented machine-learning equilibrium reconstruction with EFIT Green's functions by representing the toroidal current density in a PCA (principal component analysis) basis and constructing consistent basis functions for $\psi$ and synthetic magnetic diagnostics~\cite{mcclenaghan2024augmenting}. These works demonstrate the potential of ML for rapid equilibrium reconstruction and fast surrogate modeling. However, most of them address the inverse problem of reconstruction from measurements, while the forward Eqnet model provides a closely related fully connected neural-network surrogate for EFIT-based free-boundary flux-surface prediction. Reinforcement-learning approaches have produced striking demonstrations of magnetic control at TCV \cite{GALPERTI2024114640} and KSTAR \cite{degrave2022magnetic,seo2021feedforward}. Such approaches, and control-optimization methods more generally, benefit from fast and repeatedly queryable inner-loop models whose latency is both low and predictable.

Operator learning provides a natural framework for this requirement. Rather than solving one partial differential equation (PDE) instance at a time, a neural operator learns the map between families of inputs and their corresponding solution fields \cite{kossaifi2026librarylearningneuraloperators,duruisseaux2025guide,kovachki2023neuraloperator}. Once trained, the same model can be applied to new members of the learned family at the cost of a single forward evaluation. DeepONet established a practical architecture for nonlinear operator approximation \cite{lu2021deeponet}, while the Fourier Neural Operator (FNO) introduced spectral convolutions that have proven effective across broad classes of parametric PDEs \cite{li2021fno}. The FNO architecture is particularly well suited for GS equilibria since elliptic problems are globally coupled. A local change in current, pressure, or boundary geometry redistributes flux throughout the computational domain, and spectral convolutions provide an efficient mechanism for capturing such long-range interactions. FNOs have shown strong performance in physics applications \cite{li2023geom_fno,10.1145/3648506} and in plasma physics for reduced turbulence models and MHD-related tasks \cite{Gopakumar2024PlasmaSurrogateFNO,Carey2025NeuralOperatorEdgeSimulations}.

The constrained free-boundary regime relevant to diverted tokamaks remains relatively unexplored from this operator-learning perspective. In this regime, X-point location, separatrix topology, and isoflux consistency must be preserved simultaneously. Prior neural surrogates for GS-adjacent tasks have focused primarily on fixed-boundary formulations, diagnostic reconstruction from measurements, EFIT-based forward flux-surface prediction, or turbulence and MHD dynamics rather than geometry-conditioned neural-operator surrogates for constrained free-boundary equilibrium generation \cite{lu2023east,Joung2023GSDeepNet,wai2022neural,jang2024gs_pinn,Gopakumar2024PlasmaSurrogateFNO,Carey2025NeuralOperatorEdgeSimulations}.

In this work, we use a geometry-conditioned FNO to learn a computationally efficient representation of a constrained forward map from spatial coordinates, scalar operating parameters, and prescribed X-point coordinates to the full poloidal-flux field. This approach differs from diagnostic reconstruction models, which infer equilibria from measurements, and from actuator-conditioned forward models such as Eqnet, which use coil, vessel, and profile information to predict EFIT-based flux surfaces. The learned FNO here should therefore be regarded as a surrogate for a constrained free-boundary design problem, not as an unconstrained actuator-to-equilibrium map. In particular, we train the model on a controlled family of constrained double-null free-boundary equilibria generated with the
\textsc{FreeGS} code \cite{freegs_code} for a single fixed machine geometry and prescribed topology. The explicit conditioning on X-point coordinates is a deliberate design choice -- it anchors the divertor topology directly and is appropriate for design and control workflows in which the target X-point geometry is prescribed or measurable.

We assess the trained FNO model using both field-level and geometry-aware diagnostics. Specifically, in addition to relative $L^2$ error, we evaluate separatrix shape, X- and O-point localization, enclosed separatrix area, and boundary-flux consistency. We also test whether the predicted fields satisfy an external finite-difference GS residual diagnostic at the same level as the ground-truth \textsc{FreeGS} fields, using the same diagnostic on both predicted and reference fields. These results demonstrate that, within the prescribed topology and machine geometry studied in this work, neural operators can provide accurate, geometrically faithful, and millisecond-scale surrogates for repeated constrained free-boundary equilibrium evaluation. 

The paper is organized as follows. After the introductory discussion in this section, Section~\ref{sec:methods} describes the theoretical framework, data generation procedure, geometry-conditioned FNO architecture and training, and evaluation metrics. Section~\ref{sec:results} presents our results, namely the data-scaling study, field-level accuracy, geometry-aware validation, inference-latency benchmarks, and finite-difference GS residual diagnostic. Section~\ref{sec:discussion} discusses the implications for control and integrated modeling, interprets the observed error structure, and summarizes the main limitations and future extensions. We conclude in Section~\ref{sec:summary} with a summary of the main results and a brief outlook.

\section{Methods}
\label{sec:methods}

\subsection{Free-boundary GS formulation and data generation}
\label{sec:methods_data}

In axisymmetric geometry, ideal MHD equilibrium satisfies the
Grad--Shafranov equation for the poloidal magnetic flux $\psi(R,Z)$
\cite{grad1958hydromagnetic,wesson2011tokamaks}:
\begin{equation}
  \Delta^\star \psi(R,Z)
  = -\mu_0 R^2 \frac{dp}{d\psi}
  - \frac{1}{2}\frac{dF^2}{d\psi},
  \label{eq:gs}
\end{equation}
where $\Delta^\star = R\partial_R(R^{-1}\partial_R) + \partial_Z^2$,
$p(\psi)$ is the plasma pressure, and $F(\psi)=RB_\phi$ denotes the
toroidal-field function. For a free-boundary equilibrium, the plasma
boundary is not prescribed a priori. Instead, it must be determined
self-consistently together with the surrounding coil response and the
plasma profiles.

\begin{figure}[t!]
\centering
\includegraphics[width=0.72\linewidth]{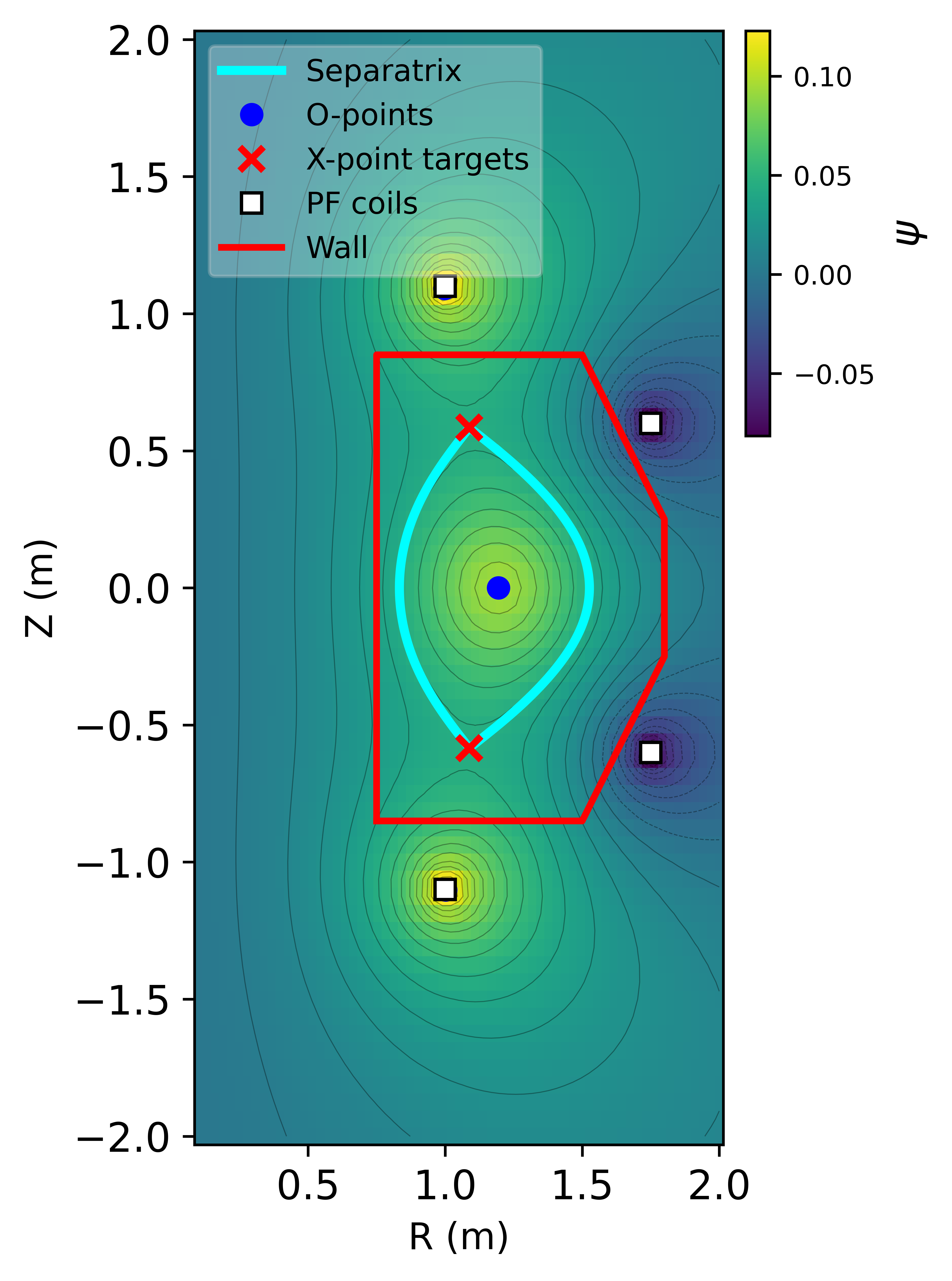}
\caption{Representative double-null free-boundary equilibrium generated with
\textsc{FreeGS} for input parameters
$P_{\mathrm{axis}} = 1200$~Pa, $I_p = 2.0\times10^{5}$~A, $f_{\mathrm{vac}} = 1.75$,
with X-point targets at $(R,Z) = (1.1,\,+0.6)$~m (upper) and $(1.1,\,-0.6)$~m (lower).
The cyan contour shows the separatrix $\psi = \psi_{\mathrm{bndry}}$, the filled blue circle is the magnetic O-point, 
red crosses mark the X-point target locations, the red polygon is the wall boundary, and open squares denote PF coil positions.
This configuration is representative of the training distribution.}
\label{fig:setup}
\end{figure}

In this study, we consider diverted double-null equilibrium configurations
constrained by prescribed upper and lower X-point locations together with an
isoflux condition. The boundary flux is defined as
$\psi_{\mathrm{bndry}} = \tfrac{1}{2}\left(\psi_X^{\mathrm{lo}} +
\psi_X^{\mathrm{up}}\right)$, where the superscripts denote the lower and
upper X-points. Each baseline equilibrium evaluation is therefore the solution
of a nonlinear elliptic PDE coupled to global constraint updates. In practice,
the problem is solved with Picard iteration while enforcing the X-point and
isoflux conditions at each step. The surrogate developed below should
therefore be interpreted as a fast approximation to this constrained
free-boundary design problem, not as an unconstrained actuator-to-equilibrium
map.

All data were generated using \textsc{FreeGS} \cite{freegs_code}, an
open-source free-boundary Grad--Shafranov solver, for the fixed \textsc{FreeGS} 
\texttt{TestTokamak} geometry. The equilibria were computed over the radial domain
$R \in [0.1, 2.0]$~m and vertical range $Z \in [-2.0, 2.0]$~m on a uniform
$65{\times}65$ grid. The pressure and toroidal-field profile shapes are fixed by
the \textsc{FreeGS} setup used here, while their amplitudes and the global
equilibrium are varied through scalar operating parameters. Thus, the dataset
represents a controlled free-boundary equilibrium family for a single machine
geometry and prescribed double-null topology, rather than a cross-device or
cross-topology distribution.

For each sample, the scalar operating parameters
$(P_{\mathrm{axis}}, I_p, f_{\mathrm{vac}})$ are drawn uniformly from
\begin{align}
  P_{\mathrm{axis}} &\in [200,\, 3000]\ \text{Pa},\\
  I_p               &\in [5{\times}10^4,\, 4{\times}10^5]\ \text{A},\\
  f_{\mathrm{vac}}  &\in [0.5,\, 3.0].
\end{align}
The reference X-point targets are $(R_X, Z_X) = (1.1, \pm 0.6)$~m.
Symmetric geometric jitter $|\delta R|, |\delta Z| \leq 0.02$~m is applied
to the upper and lower nulls in a manner that preserves exact up--down
symmetry. This broadens the separatrix family while preserving the double-null
topology.

A base training pool of 5000 equilibria is generated with random seed 123.
The scaling study uses nested subsets
$N_{\mathrm{train}} \in \{500, 1000, 2000, 5000\}$ drawn from this pool
with a fixed permutation, using seed 12345, so that larger training sets
strictly contain the smaller ones. Validation and test datasets, each
containing 500 samples, are generated independently using seeds 456 and 789,
respectively, and are held fixed across all experiments. During generation,
each candidate solution is retained only if \textsc{FreeGS} converges, the solved
plasma current satisfies
$|I_{p,\mathrm{sol}} - I_{p,\mathrm{tgt}}| / |I_{p,\mathrm{tgt}}| \leq
10\%$, and \texttt{find\_critical} returns at least two distinct X-points
on the resulting field. All 6000 candidate equilibrium solutions across the three datasets
were accepted, giving a 100\% acceptance rate. The learned distribution is
therefore identical to the sampled prior, with no implicit parameter-space
narrowing from filtering. A representative equilibrium from the training
distribution is shown in Fig.~\ref{fig:setup}.

\subsection{Geometry-conditioned FNO surrogate}
\label{sec:methods_fno}

\begin{figure*}[t!]
\centering
\includegraphics[width=0.80\linewidth]{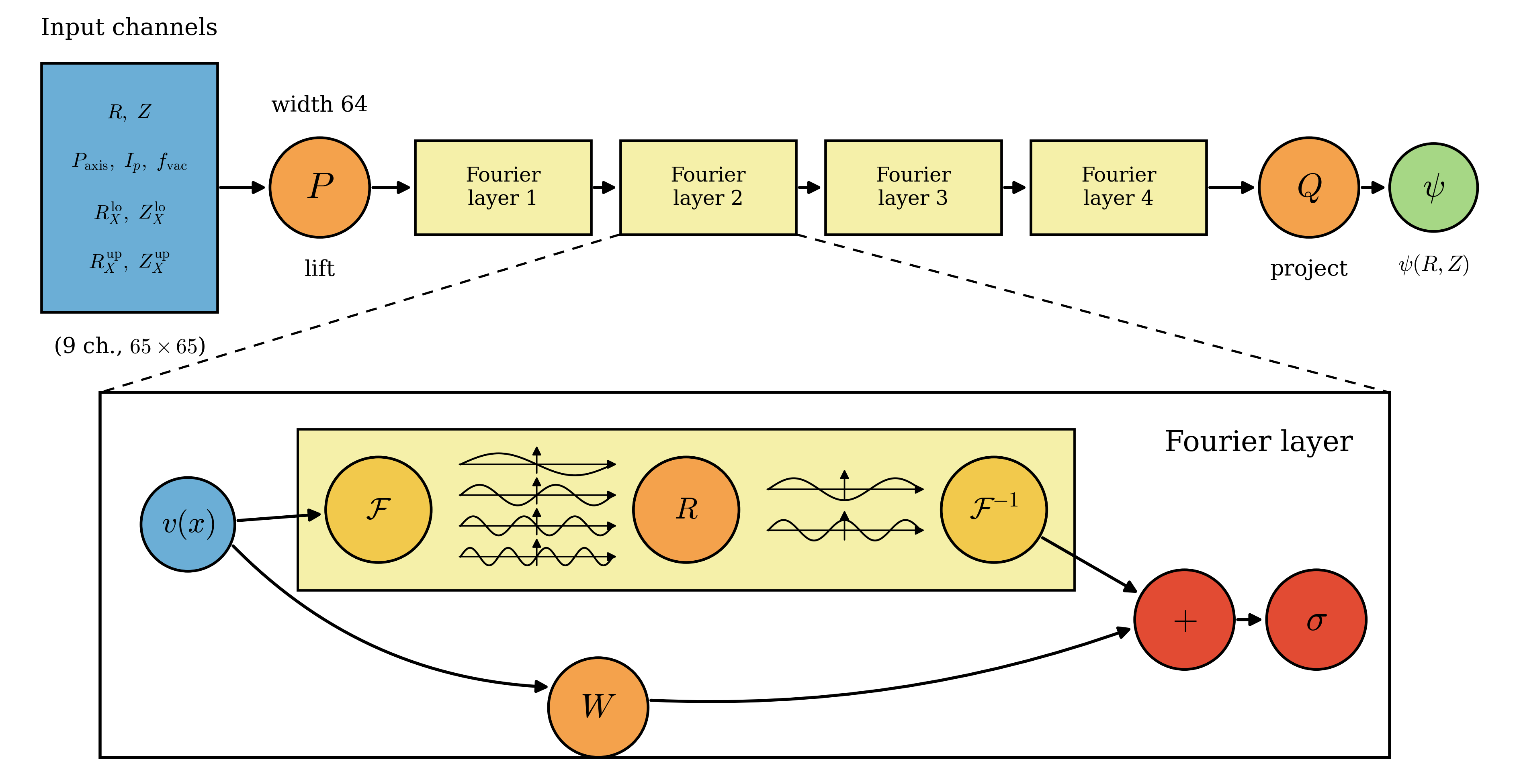}
\caption{Schematic of the geometry-conditioned Fourier Neural Operator used
in this work. The input consists of nine channels on the computational grid:
spatial coordinates $(R,Z)$, broadcast scalar operating parameters
$(P_{\mathrm{axis}},I_p,f_{\mathrm{vac}})$, and broadcast lower and upper
X-point coordinates. A lifting layer maps these channels to a hidden
representation, followed by four Fourier layers with truncated spectral
convolutions and pointwise mixing. The final projection produces the full
poloidal-flux field $\psi(R,Z)$ on a $65{\times}65$ grid. Geometry-aware
diagnostics, including separatrix shape, X-/O-point localization, and the
finite-difference Grad--Shafranov residual, are computed from the predicted
field but are not used as training targets.}
\label{fig:fno_schematic}
\end{figure*}

We approximate the parametric solution operator
\begin{equation}
  \mathcal{G}:
  \bigl(R,\,Z,\,P_{\mathrm{axis}},\,I_p,\,f_{\mathrm{vac}},\,
  R_X^{\mathrm{lo}},\,Z_X^{\mathrm{lo}},\,
  R_X^{\mathrm{up}},\,Z_X^{\mathrm{up}}\bigr)
  \mapsto \psi(R,Z),
  \label{eq:operator}
\end{equation}
which takes nine gridded input channels and maps them to the full
two-dimensional flux field on the $65{\times}65$ computational grid. The
first two channels are the spatial coordinates $(R,Z)$, which provide
explicit geometric context. The next three channels are the scalar operating
parameters $(P_{\mathrm{axis}}, I_p, f_{\mathrm{vac}})$, broadcast over the
grid. Here \(P_{\mathrm{axis}}\) is the plasma pressure at the axis, \(I_p\)
is the plasma current, and \(f_{\mathrm{vac}}\) controls the vacuum
toroidal-field contribution in the \textsc{FreeGS} setup.
The final four channels are the lower and upper X-point coordinates,
also broadcast over the grid, which specify the double-null divertor geometry
directly.

This geometry conditioning is deliberate. Rather than asking the network to
infer the separatrix topology and saddle-point placement from the scalar
operating parameters alone, we supply the X-point geometry explicitly as part
of the operator input. In effect, the learned map is conditioned both on the
operating point and on the divertor configuration to be reproduced. This
choice is appropriate for constrained design and control workflows in which
the target X-point geometry is prescribed or measurable, but it should not be
confused with an actuator-level forward model in which coil currents and
profile parameters alone determine the resulting X-point locations. 

We employ a two-dimensional Fourier Neural Operator (FNO)
\cite{li2021fno,kovachki2023neuraloperator}. For an input field
$v(\mathbf{x})$, each operator layer applies a truncated spectral
convolution,
\begin{equation}
  \mathcal{K} v(\mathbf{x})
  = \mathcal{F}^{-1}\!\left( R(\mathbf{k})\,
    \mathcal{F}[v](\mathbf{k}) \right)(\mathbf{x}),
  \label{eq:spectral_conv}
\end{equation}
where $\mathcal{F}$ denotes the discrete Fourier transform and
$R(\mathbf{k})$ is a learned complex-valued weight tensor restricted to the
lowest retained Fourier modes. Compared with purely local convolutional
kernels, this spectral representation provides efficient global mixing, which
is well matched to elliptic PDEs whose solutions depend on nonlocal coupling
across the domain.

The network contains four Fourier layers with
$n_{\mathrm{modes}} = (16,16)$ retained modes per dimension and hidden
channel width 64, for a total of 4,770,241 trainable parameters. The model is
implemented using the \texttt{neuraloperator} Python library
\cite{kossaifi2026librarylearningneuraloperators}. Spatial coordinates are
passed explicitly as input channels -- no additional positional embedding is
used. A final pointwise linear projection maps the hidden representation to
the single output channel $\psi(R,Z)$. The resulting geometry-conditioned 
FNO architecture is summarized schematically in Fig.~\ref{fig:fno_schematic}.

\subsection{Normalization and training}
\label{sec:methods_training}

The scalar inputs $(P_{\mathrm{axis}}, I_p, f_{\mathrm{vac}})$ are
standardized using the training-set mean and standard deviation. The four
X-point coordinate channels are standardized in the same way. The spatial
coordinates $(R,Z)$ are linearly mapped to $[-1,1]$. The target flux field
$\psi$ is normalized using the training-set mean and standard deviation.
The training loss and reported relative $L^2$ errors are computed in this
normalized representation. Physical-unit quantities, including the RMSE in
Wb and all geometry-aware metrics, are computed after inverse
transformation. This convention ensures that errors are evaluated
consistently across the scaling study, while the physical RMSE provides a
direct dimensional measure of the flux-field error.

Models are trained with the AdamW optimizer \cite{loshchilov2019adamw},
using an initial learning rate of $10^{-3}$ and weight decay $10^{-4}$.
The training objective is the mean squared error between predicted and target
flux fields in normalized units. A \texttt{ReduceLROnPlateau} scheduler
halves the learning rate whenever the validation relative $L^2$ error fails
to improve for 20 consecutive epochs, subject to a minimum learning rate of
$10^{-5}$. Early stopping is applied with a patience of 75 epochs, again
based on the validation relative $L^2$ error. Mini-batches contain 16
samples, and training uses 4 data-loader workers.

For each training-set size in the scaling study, three independent random
seeds are used for weight initialization. The dataset subsets themselves
are kept fixed across seeds so that the reported variability reflects model
initialization rather than data resampling.

\subsection{Evaluation metrics and timing}
\label{sec:methods_eval}

Field-level accuracy is quantified by the relative $L^2$ error,
\begin{equation}
  \varepsilon_{\mathrm{rel}} =
  \frac{\|\psi_{\mathrm{pred}} - \psi_{\mathrm{true}}\|_2}
       {\|\psi_{\mathrm{true}}\|_2},
  \label{eq:rel_l2}
\end{equation}
and by the physical RMSE in Wb. All relative $L^2$ errors reported in this
paper are computed in normalized units and then converted to percentages.
Physical RMSE values are computed after inverse-normalizing both the predicted
and target flux fields, so that they provide a direct dimensional measure of
the flux-field error.

Geometry-aware metrics are evaluated for the best $N=5000$ model over all
500 test samples. For each equilibrium, critical points of the predicted and
ground-truth fields are located with the \texttt{find\_critical} function from \textsc{FreeGS}. 
For each field, the separatrix is extracted as the largest closed contour of $\psi$ at the
mean X-point flux value,
\begin{equation}
  \psi_{\mathrm{bndry}} =
  \tfrac{1}{2}\left(\psi_X^{\mathrm{lo}} + \psi_X^{\mathrm{up}}\right).
  \label{eq:bndry_flux_eval}
\end{equation}
To avoid contaminating closed-separatrix metrics with open divertor-leg
contour artifacts, contour points whose radial distance from the contour
centroid exceeds $Q_3 + 3{\times}\mathrm{IQR}$ of the contour-point radial
distances are discarded. This post-processing is applied identically to the
true and predicted contours before distance metrics are computed. We report
the mean closest-point distance and Hausdorff distance between the predicted
and true separatrices, the relative error in the enclosed separatrix area,
the X-point location errors,
\begin{equation}
  \Delta_X^{\mathrm{lo,up}} =
  \sqrt{(R_X^{\mathrm{pred}} - R_X^{\mathrm{true}})^2
       +(Z_X^{\mathrm{pred}} - Z_X^{\mathrm{true}})^2},
  \label{eq:xpoint_error}
\end{equation}
the magnetic O-point location error, and the absolute error in
$\psi_{\mathrm{bndry}}$.

A Tukey-fence analysis, using the threshold $Q_3 + 1.5{\times}\mathrm{IQR}$,
is used to characterize the heavy tail of the X-point error distribution. On
the 500 held-out test predictions, \texttt{find\_critical} returns two
distinct X-points for every sample, for both the ground-truth \textsc{FreeGS}
fields and the FNO predictions; there are no detection failures. Nevertheless,
$5.8\%$ of samples, corresponding to $29/500$ cases, lie beyond the Tukey
threshold for at least one of the two nulls. These geometric outliers are
retained in all reported statistics and are the dominant contributor to the
heavy Hausdorff tail discussed in Sec.~\ref{sec:results_error_dist}; they are
not a distinct failure mode, but rather the upper tail of the field-level
error distribution amplified geometrically at the saddle points.

Inference latency is measured at batch size 1 with synchronized CUDA events
on an NVIDIA A100 SXM4-40GB for GPU inference. CPU inference is measured with
\texttt{time.perf\_counter} using a single PyTorch thread. Baseline CPU timing
is recorded for the iterative \textsc{FreeGS} solver using the same solver
configuration as in data generation. Median and p95 statistics are reported
over 500 test evaluations. The resulting timing comparison is therefore a
comparison against \textsc{FreeGS} as configured in these experiments;
optimized production equilibrium solvers or warm-started iterative solves
could change the absolute speedup factors.

To quantify physics consistency beyond agreement with \textsc{FreeGS} output,
we evaluate an external finite-difference Grad--Shafranov residual diagnostic
on the predicted fields. For each predicted field, we first inverse-normalize
$\psi_{\mathrm{pred}}$ and then compute $\Delta^\star \psi_{\mathrm{pred}}$
numerically using second-order central finite differences on the
$65{\times}65$ grid, where
$\Delta^\star = R\,\partial_R(R^{-1}\partial_R) + \partial_Z^2$.
The ground-truth right-hand side,
\begin{equation}
  \mathrm{RHS}_{\mathrm{true}}
  = -\mu_0 R^2\,\frac{dp}{d\psi}
    - F\,\frac{dF}{d\psi},
  \label{eq:gs_rhs_eval}
\end{equation}
is loaded directly from the pre-computed \texttt{dpdpsi} and
\texttt{FdFdpsi} fields stored in the dataset, evaluated on the corresponding
ground-truth equilibrium. The per-sample normalized residual is
\begin{equation}
  \mathcal{R}_i =
  \frac{\|\Delta^\star\psi_{\mathrm{pred}} -
        \mathrm{RHS}_{\mathrm{true}}\|_{2,\Omega_i}}
       {\|\mathrm{RHS}_{\mathrm{true}}\|_{2,\Omega_i}},
  \label{eq:gs_residual}
\end{equation}
where $\|\cdot\|_{2,\Omega_i}$ denotes the $L^2$ norm restricted to interior
grid points inside the plasma mask $\Omega_i$. The same quantity is computed
for the ground-truth \textsc{FreeGS} fields to establish the finite-difference
baseline. The absolute value of this diagnostic should therefore not be
interpreted as a solver-internal residual norm; it is a comparative
finite-difference diagnostic applied consistently to both predicted and
reference fields.

\section{Results}
\label{sec:results}

\subsection{Field accuracy and data scaling}
\label{sec:results_field_scaling}

\begin{table}[t!]
\centering
\caption{Field-level accuracy versus training set size.
         Values are averaged over three random
         initializations. Relative $L^2$ errors are computed in
         normalized units and reported as percentages. Physical RMSE is
         reported in units of Wb after inverse normalization.}
\label{tab:field_accuracy}
\begin{tabular}{lcccc}
\toprule
$N_{\mathrm{train}}$ & rel$L^2$ (\%) & std (\%) &
  RMSE$_{\mathrm{norm}}$ & RMSE$_{\mathrm{phys}}$ (Wb) \\
\midrule
  500 & 0.286 & 0.034 & $3.13\times10^{-3}$ & $8.53\times10^{-5}$ \\
 1000 & 0.182 & 0.002 & $1.97\times10^{-3}$ & $5.33\times10^{-5}$ \\
 2000 & 0.109 & 0.004 & $1.18\times10^{-3}$ & $3.16\times10^{-5}$ \\
 5000 & 0.061 & 0.006 & $6.59\times10^{-4}$ & $1.79\times10^{-5}$ \\
\bottomrule
\end{tabular}
\end{table}

The most global quantitative trend in the study is the monotonic reduction
in test error with increasing training data. Table~\ref{tab:field_accuracy}
reports the test relative $L^2$ error and physical RMSE for each
$N_{\mathrm{train}}$, averaged over three random initializations.

\begin{figure}[t!]
\centering
\includegraphics[width=\linewidth]{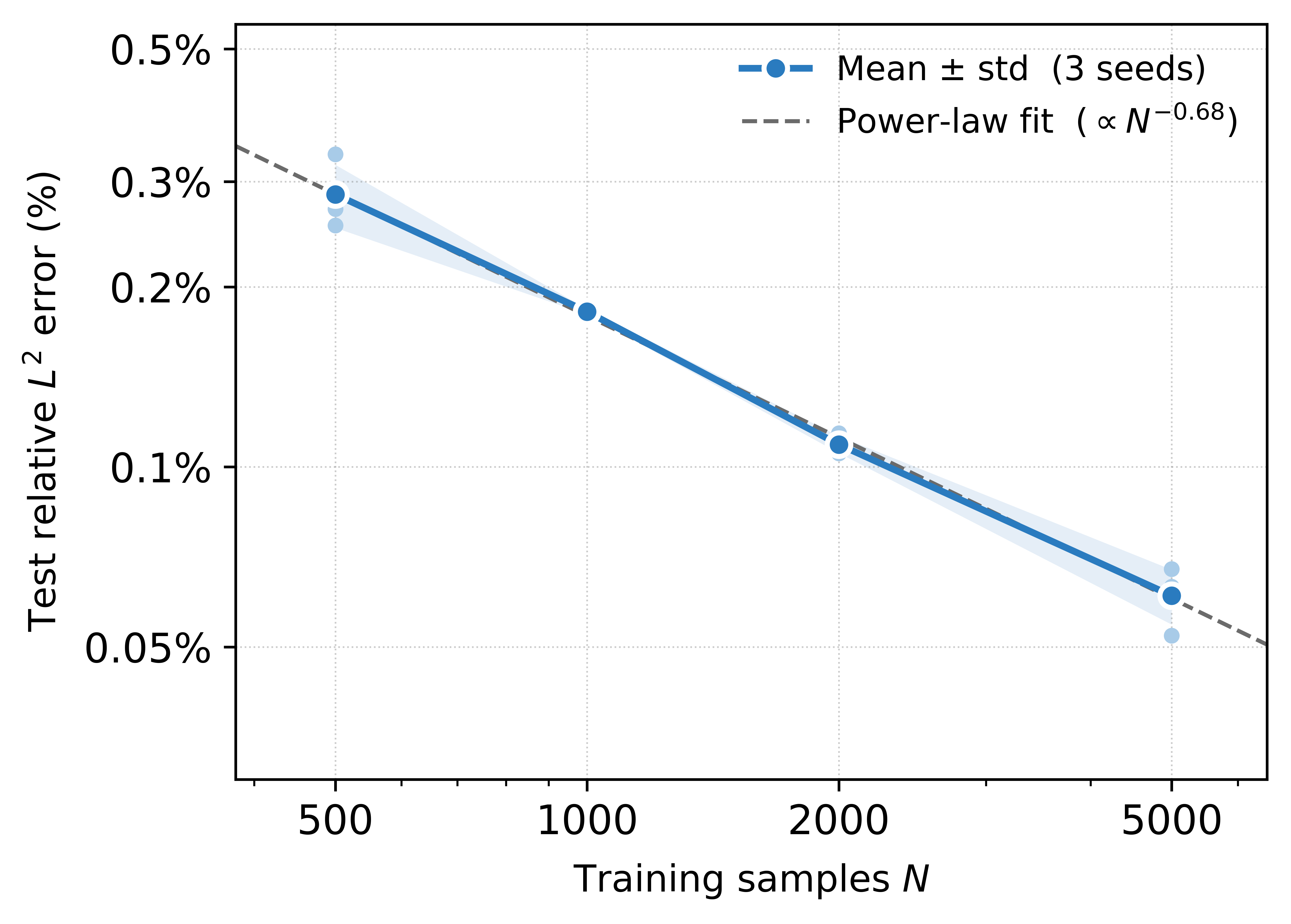}
\caption{
Test relative $L^2$ error versus number of training samples on a log-log
scale. Filled circles denote means over three initialization seeds, the
shaded band spans $\pm1$ standard deviation and the small open markers show
individual seed results. The dashed line is a power-law fit
$\varepsilon \propto N^{-0.68}$ over the tested range. The fitted trend is
consistent with the model exploiting the regularity of the constrained
equilibrium family, but is not intended as a definitive sample-complexity
law.}
\label{fig:scaling_curve}
\end{figure}

Figure~\ref{fig:scaling_curve} shows the same trend on a log-log scale. A
power-law fit to the mean errors yields
\begin{equation}
  \varepsilon_{\mathrm{rel}} \propto N^{-0.68},
\end{equation}
which is steeper than the $N^{-0.5}$ behavior expected from random
sampling alone. Fit over four training sizes, this empirical power-law
is suggestive rather than definitive -- it is consistent with the model
exploiting the regularity of the constrained equilibrium family within
the tested configuration space, but we do not claim it as a precise
characterization of the operator's sample complexity. Across the full
range from 500 to 5000 training samples, the mean test error drops by
a factor of 4.7. By $N=1000$, the seed-to-seed variance is already very
small (std $= 0.002\%$), indicating a stable training regime.

The best-performing model ($N=5000$, seed 3, best validation epoch 310)
achieves a mean relative $L^2$ error of $0.052\%$ and a median of
$0.046\%$ on the 500-sample test dataset. The corresponding physical
RMSE is $1.54\times10^{-5}$~Wb. More than 95\% of test samples fall below
$0.12\%$ relative error, so the reported mean is representative of the bulk
of the distribution rather than being driven by a few favorable cases. The
full distributions of field-level and geometry-aware errors are shown in
Fig.~\ref{fig:error_distributions}.

\subsection{Geometry-aware validation}
\label{sec:results_geometry}

Field norms alone are not sufficient for diverted equilibria. Additionally, a key question is whether the learned flux preserves the separatrix and the associated critical-point structure.  We identify the X-points as saddle points satisfying $\nabla \psi = 0$, while the separatrix is defined by $\psi = \psi_{\mathrm{bndry}}$, where $\psi_{\mathrm{bndry}} = \tfrac{1}{2}(\psi_X^{\mathrm{lo}} + \psi_X^{\mathrm{up}})$. The geometry-aware metrics for the best \(N = 5000\) model across all
500 test equilibria are summarized in Table~\ref{tab:geometry_metrics}.

\begin{table}[t!]
\centering
\caption{Geometry-aware metrics for the best model ($N=5000$, seed 3) over
         500 test equilibria. All distance metrics are in cm.
         \texttt{find\_critical} returns two distinct X-points on all 500
         test samples; $5.8\%$ of samples ($29/500$) lie beyond the
         Tukey fence ($Q_3 + 1.5{\times}\mathrm{IQR}$) for at least one
         X-point. All samples are included in the reported statistics.}
\label{tab:geometry_metrics}
\begin{tabular}{lccc}
\toprule
Metric & Mean & Median & P95 \\
\midrule
Separatrix mean error (cm)   & 0.072 & 0.059 & 0.147 \\
Separatrix Hausdorff (cm)    & 2.128 & 0.967 & 4.693 \\
Separatrix area rel.\ error (\%) & 0.465 & 0.343 & 0.992 \\
X-point lower (cm)           & 0.161 & 0.136 & 0.350 \\
X-point upper (cm)           & 0.112 & 0.097 & 0.236 \\
O-point (cm)                 & 0.031 & 0.023 & 0.079 \\
$|\Delta\psi_{\mathrm{bndry}}|$ (Wb) & $9.82\times10^{-6}$ & --- & $7.00\times10^{-5}$ \\
\bottomrule
\end{tabular}
\end{table}

The separatrix is reproduced with a mean closest-point deviation of 0.072~cm and a median Hausdorff distance of 0.967~cm, while the enclosed
area error remains below 0.5\% on average. The mean Hausdorff distance is larger (2.128~cm) because it is sensitive to a small subset of cusp-region
outliers near the X-point pinch, as discussed below. Both X-points are recovered to within approximately 0.1--0.2~cm on average, with the upper
null slightly more accurate than the lower one. This residual gap is examined further in Sec.~\ref{sec:discussion_implications}. The magnetic O-point
is localized to within 0.031~cm (about 0.3~mm), and the mean boundary-flux error is $9.82\times10^{-6}$~Wb.

\begin{figure}[t!]
\centering
\includegraphics[width=0.75\linewidth]{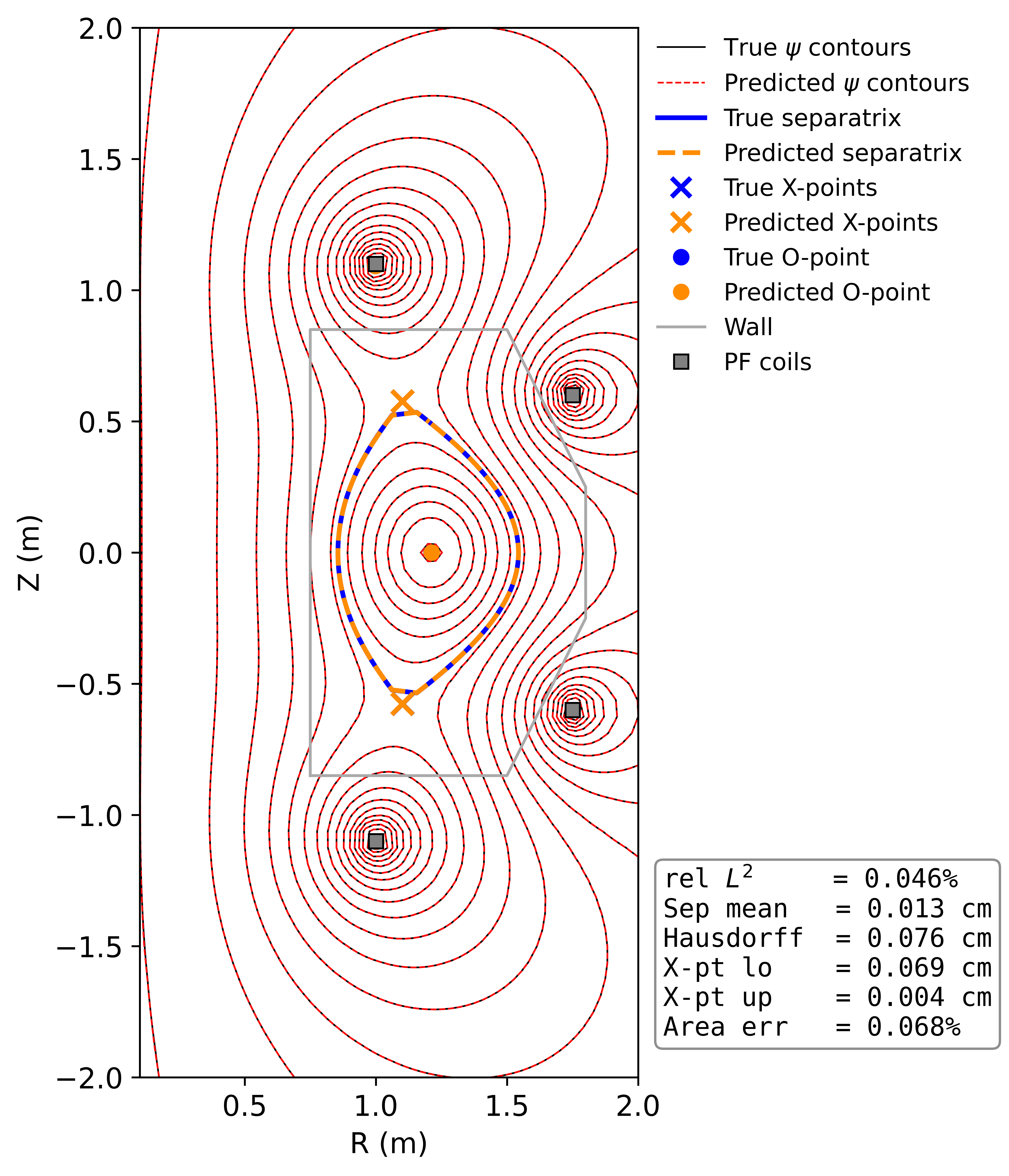}
\caption{Poloidal flux contour overlay for a near-median test equilibrium
(rel.\ $L^2 = 0.046\%$,
$P_{\mathrm{axis}} = 874$~Pa,
$I_p = 1.53\times10^{5}$~A,
$f_{\mathrm{vac}} = 2.11$).
Solid black lines are ground-truth $\psi$ contours and dashed red lines are
FNO predictions on identical iso-flux levels. The true separatrix
($\psi = \psi_{\mathrm{bndry}}$) is shown in dark blue (solid) and the
predicted separatrix in orange (dashed). The true and predicted X-points are
marked with blue and orange crosses, and the O-points with filled
circles of the same colors, respectively. The gray polygon is the wall boundary and the gray
squares are PF coil positions. The metrics box reports field-level and
geometry-aware errors for this single sample, not ensemble statistics --
ensemble means and medians are given in Table~\ref{tab:geometry_metrics}.
The residual discrepancies near the X-points are consistent with the
geometric amplification of small flux errors at saddle points of $\psi$.}
\label{fig:contour_overlay}
\end{figure}

Figure~\ref{fig:contour_overlay} shows a near-median test case in which the predicted and reference contour families almost overlap. The near-perfect
overlap of the flux contours across the domain illustrates that the low relative $L^2$ error corresponds to a geometrically faithful equilibrium, while the remaining visible discrepancies are localized near the X-points.

\subsection{Error structure and separatrix ensemble behavior}
\label{sec:results_error_dist}

\begin{figure*}[t!]
\centering
\includegraphics[width=0.52\textwidth]{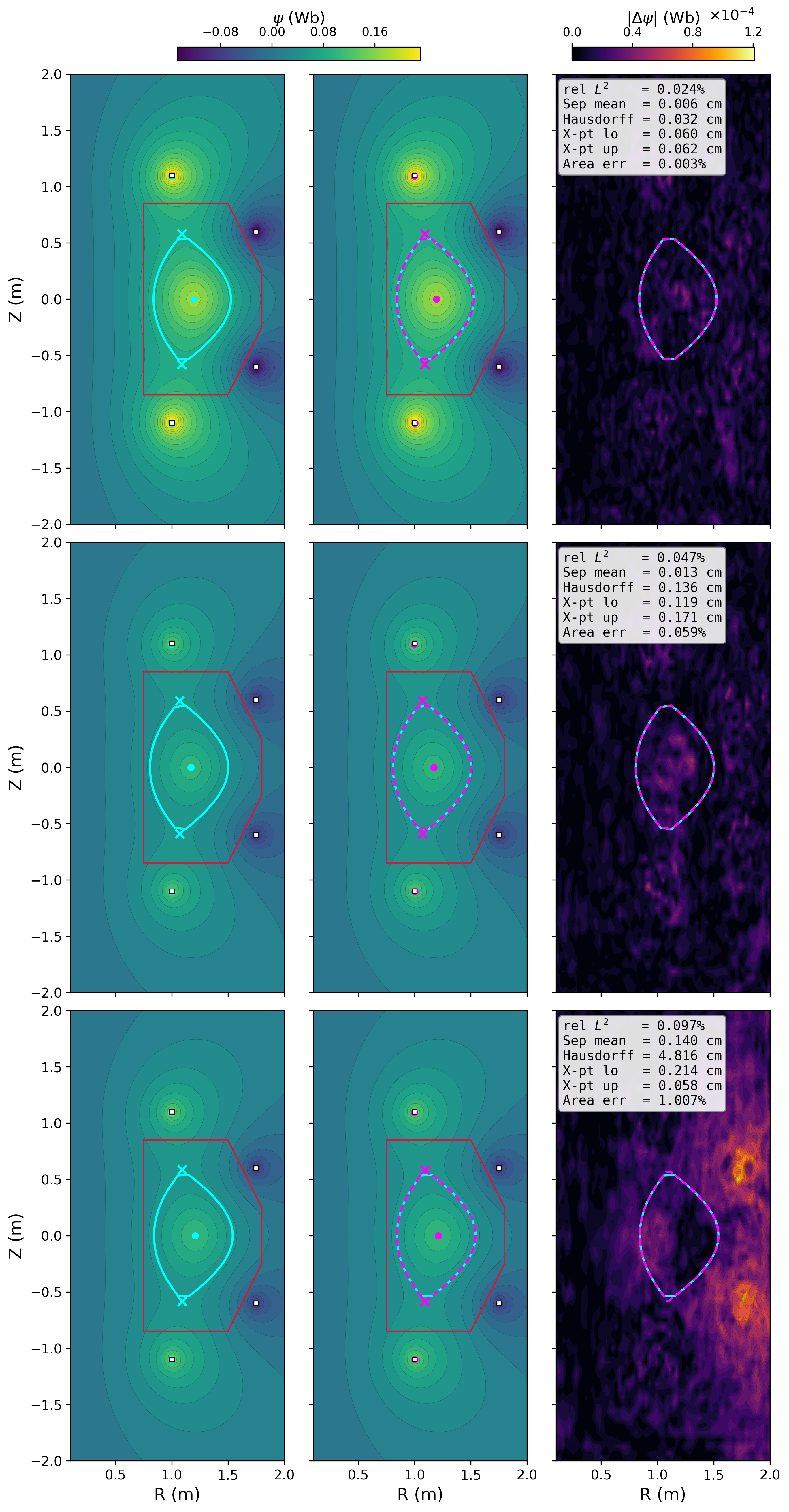}
\caption{Poloidal flux field reconstructions for three representative test
equilibria selected by relative $L^2$ error: low error (top row,
$\varepsilon_{\rm rel} = 0.024\%$), near-median error (middle row,
$\varepsilon_{\rm rel} = 0.047\%$), and near-p95 error (bottom row,
$\varepsilon_{\rm rel} = 0.097\%$).
Each row shows, from left to right: the ground-truth $\psi(R,Z)$, the FNO
prediction, and the absolute pointwise error
$|\psi_{\mathrm{pred}} - \psi_{\mathrm{true}}|$.
The two field panels in each row share a common color scale (top
colorbar), while the error panel uses a separate scale to resolve spatial error
structure across the full dynamic range. In the true $\psi$ panels, the cyan contour 
marks the ground-truth separatrix. In the predicted and error panels, the true separatrix is
shown in cyan (solid) and the predicted separatrix in magenta (dashed).
The gray polygon denotes the wall boundary and open squares mark PF coil
positions. Pointwise error is spatially concentrated near the X-points and along the
separatrix, with core-region errors remaining uniformly small across all
three cases. Per-panel geometry-aware metrics are reported in the inset boxes.}
\label{fig:snapshots}
\end{figure*}

Figure~\ref{fig:snapshots} presents three representative test equilibria: a low-error case, a near-median case, and a near-p95 case. Across all three examples, the global flux-surface structure is reproduced faithfully. The predicted equilibria preserve the separatrix topology and X-point geometry, and no spurious islands or discontinuities are observed.

In addition, it is observed that the error field is smooth and spatially localized. It concentrates near the high-curvature regions around the X-points and along the separatrix, where small perturbations in $\psi$ have the largest geometric effect. Core-region errors remain uniformly small. Even for near-p95 samples, the
dominant discrepancy is a modest local shift in flux magnitude rather than a structural failure of the reconstructed equilibrium.

\begin{figure*}[t!]
\centering
\includegraphics[width=0.75\textwidth]{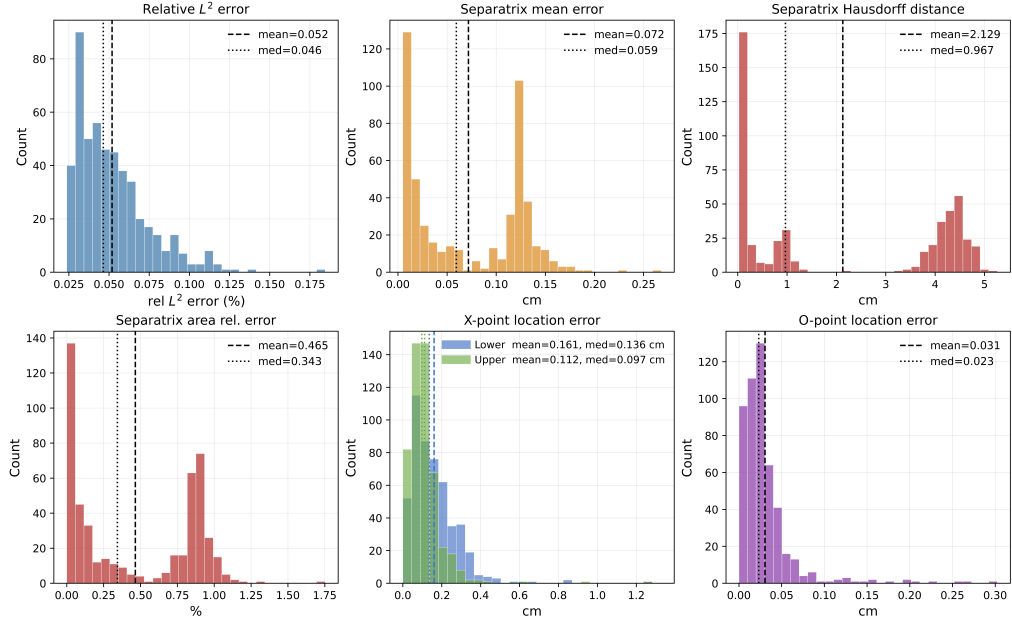}
\caption{Distributions of field-level and geometry-aware errors over all
500 test equilibria for the best model ($N=5000$, seed~3).
Top row, left to right: relative $L^2$ error, separatrix mean
closest-point deviation, and separatrix Hausdorff distance.
Bottom row: separatrix enclosed area relative error, X-point location
errors for the lower (blue) and upper (green) X-points shown as
overlapping histograms, and magnetic O-point location error.
Dashed and dotted vertical lines mark the mean and median of each
distribution, respectively.
The heavy right tail in the Hausdorff distribution is driven by a small
subset of samples where large deviations arise near the X-point pinch
geometry. It does not reflect systematic shape misrepresentation across
the ensemble; \texttt{find\_critical} succeeds on all 500 test
predictions, and the Tukey-outlier samples correspond to the upper tail
of the field-level error distribution amplified geometrically at the
saddle (see Sec.~\ref{sec:results_error_dist}).
The upper X-point is recovered with systematically lower error than the
lower X-point (mean $0.112$ vs $0.161$~cm).}
\label{fig:error_distributions}
\end{figure*}

Figure~\ref{fig:error_distributions} shows the distributions of field-level
and geometry-aware errors over all 500 held-out test equilibria. The
relative $L^2$ error distribution is unimodal and right-skewed, with mean
$0.052\%$ and median $0.046\%$, confirming that the best-model performance
is typical of the test set rather than exceptional.

The separatrix mean-error distribution is similarly concentrated, with most
samples below 0.10~cm. By contrast, the Hausdorff distribution is bimodal:
a dominant peak lies below 1~cm, while a secondary peak near 4--5~cm comes
from the small subset of cases in which the X-point pinch geometry
produces a large local deviation. \texttt{find\_critical} succeeds on
all 500 test predictions, so the tail is geometric rather than a detection
failure. The X-point histogram also reveals a systematic asymmetry, with
the upper X-point recovered more accurately than the lower one (mean
0.112 vs 0.161~cm). The origin of this gap is characterized further in
Sec.~\ref{sec:discussion_implications}. O-point errors are tightly concentrated
below 0.05~cm for the great majority of samples.

\textit{Unified characterization of the tail.}
We verified that the elevated X-point error in the upper tail is not a
distinct failure mode by comparing the 29 Tukey-outlier samples against
the 471 inliers. The outlier group exhibits relative $L^2$ error of
$0.075\%$ versus $0.050\%$ for inliers ($1.50\times$ ratio), O-point
error of $0.048$~cm versus $0.029$~cm ($1.63\times$), and separatrix
mean error of $0.092$~cm versus $0.071$~cm ($1.30\times$). All three
metrics degrade in concert, with the strongest statistical association
seen for the field-level error itself (point-biserial correlation
$r = 0.26$, $p = 3{\times}10^{-9}$). The heavy Hausdorff tail is therefore
a single underlying phenomenon -- samples with slightly lower field-level
accuracy produce proportionally larger geometric error, amplified most
strongly at the X-point saddle, where local contour location is
first-order sensitive to $\psi$.

\begin{figure}[t!]
\centering
\includegraphics[width=0.75\linewidth]{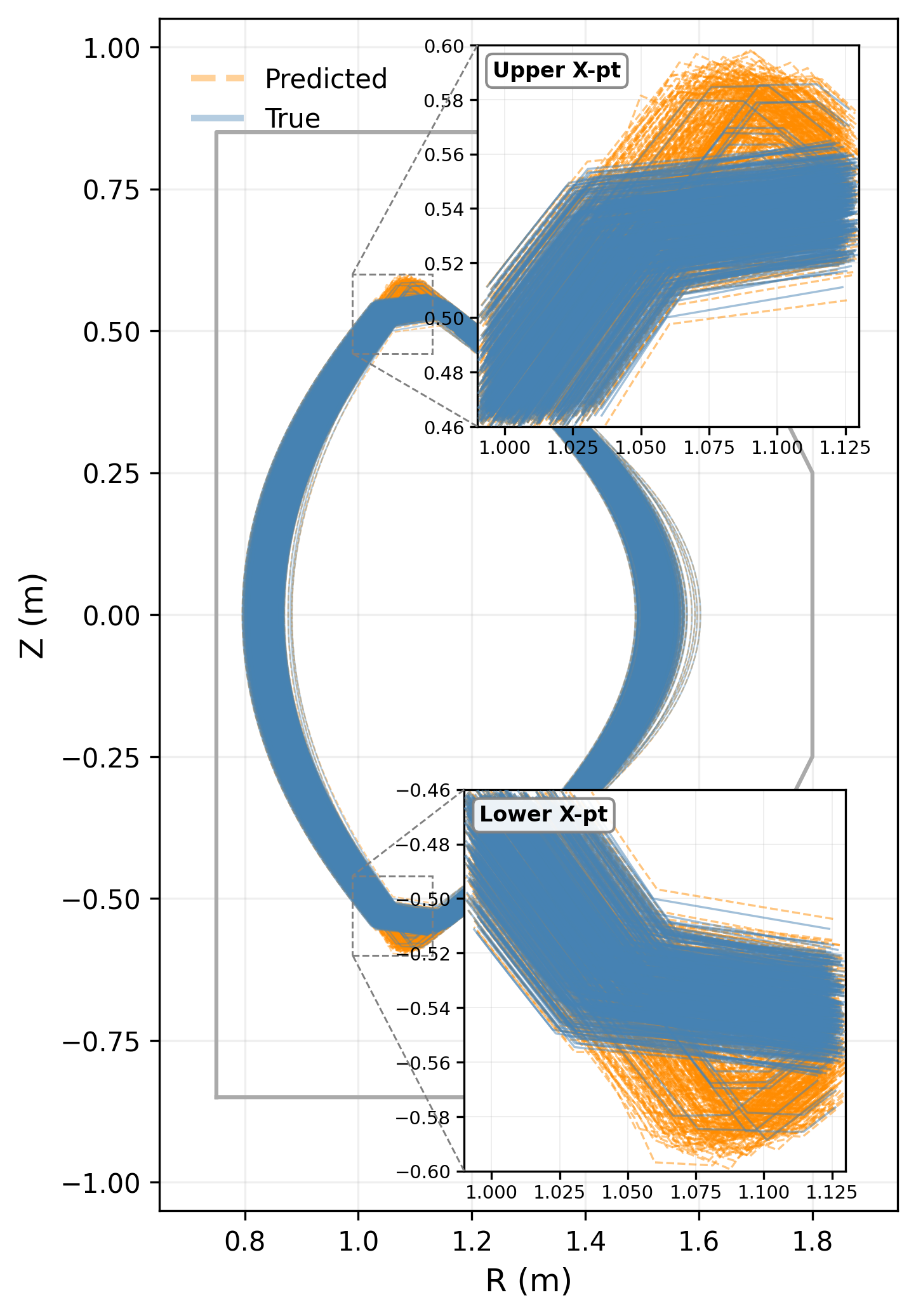}
\caption{Overlay of all 500 true (blue, solid) and predicted (orange,
dashed) separatrices from the held-out test set.
True and predicted separatrices are in close agreement across the full
test distribution, with the predicted bundle nearly indistinguishable
from the true bundle along the main arc.
Insets show zoomed views of the upper and lower X-point regions, where
localized discrepancies are visible as the separatrix pinches to a cusp.
These deviations are responsible for the elevated mean Hausdorff distance
relative to the median ($2.128$ vs $0.967$~cm) and reflect the geometric
amplification of small flux errors near saddle points of $\psi$, rather
than systematic shape misrepresentation across the separatrix.
The wall boundary is shown in gray. PF coil positions are omitted for
clarity.}
\label{fig:separatrix_overlay}
\end{figure}

Figure~\ref{fig:separatrix_overlay} overlays all 500 true and predicted
separatrices from the held-out test set. Along the main arc, the two
bundles are nearly indistinguishable. Visible spread is confined almost
entirely to the zoomed regions around the upper and lower X-points, where
the separatrix pinches to a cusp and small field errors are geometrically
amplified.

This ensemble view clarifies the interpretation of the geometry metrics. In
particular, the elevated mean Hausdorff distance is not evidence of a
systematic global shape mismatch; rather, it is driven by localized cusp
errors in a minority of samples. That behavior is consistent with the
smooth error fields in Fig.~\ref{fig:snapshots} and with the concentrated
separatrix area and mean-distance statistics in
Table~\ref{tab:geometry_metrics}.

\subsection{Computational performance and finite-difference GS residual}
\label{sec:results_timing_residual}

Inference latency is evaluated at batch size~1 using two timing methods:
synchronized CUDA events on an NVIDIA A100 SXM4-40GB for GPU inference,
and \texttt{time.perf\_counter} for CPU inference (one PyTorch thread).
\textsc{FreeGS} baseline timing is recorded on the same CPU using the
identical solver configuration used during data generation: Picard
iteration, $\gamma=10^{-12}$, isoflux constraints to a fixed outboard
midplane reference, and a maximum of 50 iterations. All 500 test
equilibria converged successfully in the \textsc{FreeGS} baseline run.
We note that \textsc{FreeGS} is an open-source Python research solver
rather than a production real-time equilibrium code. Comparisons
against hand-optimized C or Fortran free-boundary solvers, or against
\textsc{FreeGS} with warm-started initial guesses, would likely
narrow the absolute speedup while leaving the qualitative contrast
-- second-scale iterative computation versus millisecond-scale
deterministic inference -- unchanged. Table~\ref{tab:timing} summarizes 
latency statistics and speedup factors.

\begin{table}[t!]
\centering
\caption{Inference latency and speedup over all 500 test
         equilibria at batch size~1.
         GPU measurements use synchronized CUDA events on an
         NVIDIA A100 SXM4-40GB.
         CPU measurements use a single PyTorch thread.
         \textsc{FreeGS} baseline uses Picard iteration with the same
         solver configuration as data generation.
         All 500 samples converged successfully.}
\label{tab:timing}
\begin{tabular}{lcccc}
\toprule
Method & Median (ms) & p95 (ms) & \shortstack{p95\\(median)} & \shortstack{Speedup\\(median)} \\
\midrule
FNO (GPU)       &  2.765 &  2.803 & 1.01 & $\sim 640\times$ \\
FNO (CPU)       & 25.587 & 25.810 & 1.01 & $\sim  69\times$ \\
\textsc{FreeGS} (CPU) & 1768   & 2002   & 1.13 & ---                 \\
\bottomrule
\end{tabular}
\end{table}

The FNO surrogate reduces equilibrium evaluation time from second-scale
iterative computation to millisecond-scale inference. GPU and CPU speedups
relative to \textsc{FreeGS} as configured in these experiments are
${\sim}640\times$ and ${\sim}69\times$ at the median, respectively.
The CPU result is particularly relevant for deployment contexts where GPU
hardware is unavailable or where deterministic execution guarantees are
required without GPU scheduling overhead.

\begin{figure}[t!]
\centering
\includegraphics[width=\linewidth]{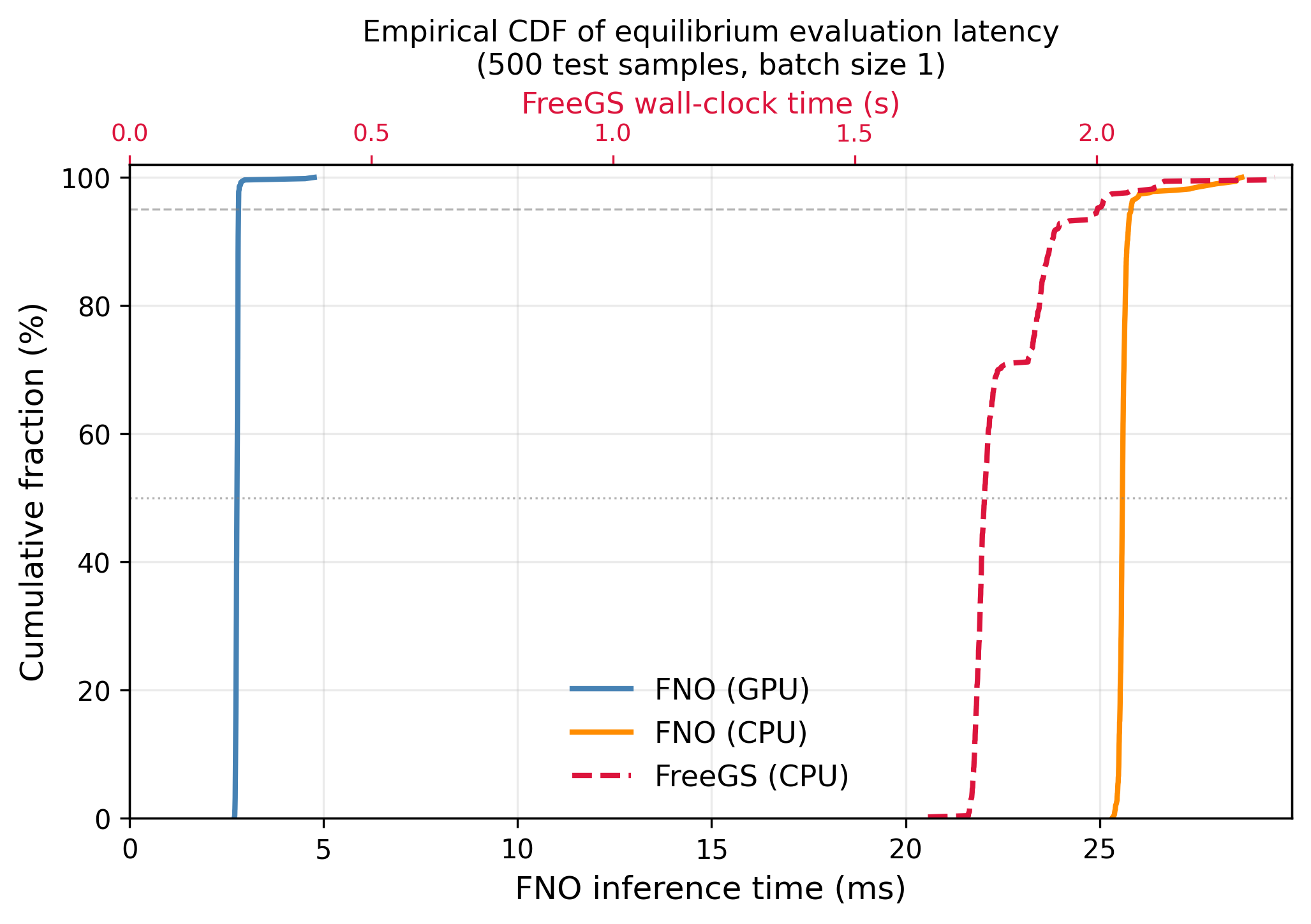}
\caption{Empirical cumulative distribution of equilibrium evaluation
latency over 500 held-out test samples at batch size~1. The FNO
surrogate on GPU (blue) and CPU (orange) exhibit near-deterministic
millisecond-scale latency (p95/median $= 1.01$ for both). The
\textsc{FreeGS} iterative solver (red dashed, top axis in seconds)
operates at second-scale cost with significant convergence-dependent
spread (p95/median $= 1.13$, std $= 92$~ms). GPU and CPU FNO speedups
are ${\sim}640\times$ and ${\sim}69\times$ at the median,
respectively, relative to \textsc{FreeGS} as configured here.}
\label{fig:latency_cdf}
\end{figure}

Equally important is the near-deterministic evaluation time of both FNO
variants. The p95/median ratio is $1.01$ for FNO (GPU) and FNO (CPU),
compared to $1.13$ for \textsc{FreeGS}, whose convergence latency varies
with the difficulty of each individual equilibrium. The standard deviation
of \textsc{FreeGS} solve times ($92$~ms) is comparable to the entire
range of FNO GPU latencies ($2.70$--$4.76$~ms across all 500 samples).
From a control-theoretic perspective, predictable evaluation latency is
as important as average speed for embedding equilibrium calls inside
real-time supervisory loops. Figure~\ref{fig:latency_cdf} shows the empirical 
CDFs (Cumulative Distribution Function) of evaluation latency for all three 
methods over the 500 test samples.

Beyond agreement with \textsc{FreeGS} output, we evaluate whether the
FNO-predicted fields are physically consistent according to the external
finite-difference Grad--Shafranov residual diagnostic defined in
Sec.~\ref{sec:methods_eval}. For each of the 500 test predictions, we
compute $\Delta^\star\psi_{\mathrm{pred}}$ by second-order central finite
differences and compare it to the ground-truth right-hand side
$\mathrm{RHS} = -\mu_0 R^2\,\mathrm{d}p/\mathrm{d}\psi
- F\,\mathrm{d}F/\mathrm{d}\psi$, evaluated inside the plasma mask.

The FNO achieves a mean normalized residual of $\mathcal{R} = 2.29$
(median $2.28$, p95 $2.40$), compared to a baseline of
$2.29 \pm 0.06$ obtained by applying the same finite-difference
scheme to the ground-truth \textsc{FreeGS} fields. The two values
are indistinguishable: the ratio of FNO to ground-truth normalized
residual is $0.998$. The non-zero baseline reflects the finite-difference
truncation error on the $65{\times}65$ grid and a slight inconsistency
between \textsc{FreeGS}'s internal discretization and our external
second-order evaluation -- both sources affect predicted and true fields
under this diagnostic. Thus, within the tested configuration space, the
surrogate satisfies the finite-difference GS residual diagnostic at the
same level as the ground-truth \textsc{FreeGS} fields. The absolute value
of $\mathcal{R}$ should not be interpreted as a solver-internal residual
norm, but as a comparative diagnostic applied consistently to both
predicted and reference fields.

\section{Discussion}
\label{sec:discussion}

\subsection{Implications and interpretation}
\label{sec:discussion_implications}

The main result of this study is not merely a reduction in field error, but a change in how constrained free-boundary equilibria can be computed in practice. Once trained, the neural operator replaces the conventional iterative nonlinear Picard scheme with a single deterministic forward evaluation. This is significant because it makes the method directly applicable to workflows in which the equilibrium solve is embedded within a larger outer loop -- including model predictive control, rapid scenario exploration, uncertainty quantification, and digital-twin architectures coupling equilibrium, actuator, and transport models. Integration into any of these specific workflows would require end-to-end validation, which we leave to future work.

Experimental real-time reconstruction codes are typically optimized for device-specific diagnostic inversion and rely on temporal coherence between consecutive time slices \cite{ferron1998rtefit,moret2015liuqe}. The surrogate presented here operates on a different principle: it maps directly from the operating point and prescribed X-point geometry to the flux field, without requiring warm starts or sequential continuity between time slices. In this respect, it should be interpreted as a surrogate for a constrained free-boundary design problem, rather than as an unconstrained actuator-to-equilibrium map. This distinction is what makes it well suited to simulation-based design and optimization tasks, where equilibria at arbitrary operating points and target geometries must be evaluated rapidly and independently of one another.

Field-level relative errors on the order of $5\times10^{-4}$ are already
sufficiently small that geometric quantities become the decisive test of
performance. It is found that the model recovers the separatrix with a
mean closest-point deviation of 0.072~cm and a median Hausdorff distance
of 0.967~cm. The X-points are located to within 0.112~cm (upper) and
0.161~cm (lower) on average, and the O-point to within 0.031~cm.
Figs.~\ref{fig:contour_overlay}--\ref{fig:separatrix_overlay} show that
the residual discrepancies concentrate near the X-point pinch, where
small perturbations in $\psi$ produce the largest geometric shifts.
Importantly, no test case exhibits a topological failure, such as a
spurious flux island or a grossly misplaced separatrix.

The persistent gap between upper and lower X-point accuracy (0.112 vs
0.161~cm) is notable given that the data-generation pipeline is exactly
up--down symmetric by construction. We tested two candidate sources of
this asymmetry directly on all 500 test samples. Applying
\texttt{find\_critical} to the ground-truth \textsc{FreeGS} fields and
comparing to the nominal symmetric targets gives mean errors of
$2.0652$~cm and $2.0651$~cm for the lower and upper nulls, respectively --
a gap of $2\times10^{-4}$~cm, consistent with numerical noise. Applying
\texttt{find\_critical} to each field and to its $Z$-reflected
counterpart gives X-point locations that agree to machine precision.
It follows that neither the solver nor the critical-point detector
contributes to the observed asymmetry. The residual $0.049$~cm gap is
most plausibly attributed to a small sensitivity to input-coordinate
conventions in a model trained without up--down data augmentation. We
note that the gap lies well below the Tukey-outlier thresholds
($0.43$~cm for the lower null, $0.25$~cm for the upper) and does not
affect any reported conclusion. Enforcing equivariance through
training-time $Z$-flip augmentation, or through an equivariant
architectural prior, would likely eliminate this residual asymmetry.

The empirical $N^{-0.68}$ power-law observed over the tested range is
steeper than the $N^{-0.5}$ rate expected from random sampling of a
generic high-dimensional function. Since this trend is fit over only
four training sizes, it should be regarded as suggestive rather than
definitive. It is consistent with the parametric equilibrium family
lying on a relatively low-dimensional effective manifold: the scalar
operating parameters and X-point coordinates define a smooth
deformation of a single double-null topology, and the GS solution
operator inherits this smoothness. Whether this behavior persists
across broader topology families or machine geometries remains to be
tested in future work.

The spatial error structure is also physically interpretable. The
pointwise error is found to concentrate near the X-point pinch geometry
and along the separatrix, while errors in the core region remain small
throughout (Fig.~\ref{fig:snapshots}). Near an X-point, the flux
gradient vanishes, so that a small absolute error in $\psi$ translates
into a comparatively large displacement of the separatrix contour. This
geometric amplification effect accounts for both the elevated mean
Hausdorff distance relative to the median ($2.128$ versus $0.967$~cm)
and the bimodal tail seen in Fig.~\ref{fig:error_distributions}, which
reflects a minority of geometrically sensitive cases rather than a
systematic misrepresentation of shape.

\subsection{Scope and limitations}
\label{sec:discussion_limitations}

This work is deliberately scoped to a single machine geometry and a
single topology, and the results should be interpreted accordingly. All
training and test data are drawn from a single wall geometry and a
single controlled double-null equilibrium family. Accordingly, the
results demonstrate strong in-distribution generalization within that
family, and should not be taken to imply cross-device or cross-topology
transfer. The operator is conditioned on prescribed X-point coordinates,
which is natural for design and control workflows in which the divertor
geometry is specified, but it does not address diagnostic reconstruction
problems in which the geometry must instead be inferred from
measurements.

The present model also does not constitute a general actuator-level
forward solver. In an actuator-conditioned formulation, the coil
currents, vessel currents, and profile information determine the
equilibrium, with the X-point locations obtained as outputs. Here, by
contrast, the X-point coordinates form part of the input specification.
This is a useful formulation for constrained scenario design and
target-geometry studies, but it should not be interpreted as a
replacement for actuator-to-equilibrium modeling.

All experiments were performed on a fixed $65{\times}65$ grid; transfer
to other resolutions was not tested. The surrogate models static GS
equilibria and is not yet coupled to transport evolution, actuator
dynamics, or stability calculations. The speedup factors reported here
are measured relative to \textsc{FreeGS} as configured in the present
experiments. Optimized production equilibrium solvers, hand-tuned C or
Fortran implementations, or warm-started iterative solves may reduce
the absolute speedup factors obtained; however, the qualitative
contrast between iterative, sample-dependent latency and deterministic,
millisecond-scale inference is expected to remain. Finally,
experimental deployment would require hardened software integration and
careful validation under distribution shift, both of which lie beyond
the scope of the present work.

\subsection{Future extensions and broader perspective}
\label{sec:discussion_future}

A natural next step is the construction of coupled equilibrium--transport
operator surrogates. In integrated modeling, transport solvers repeatedly
call equilibrium routines to update the magnetic geometry, and the
repeated GS solve is often the dominant cost
\cite{meneghini2017selfconsistent}. A paired surrogate could thus enable
fast self-consistent updates within optimization and digital-twin loops.

A second direction is to broaden the training distribution across
magnetic topologies, including lower and upper single-null, limiter, and
snowflake configurations, and eventually across machine geometries. This
would be valuable for cross-device studies and for controller design on
next-step devices such as ITER~\cite{Shimada_2007} and
SPARC~\cite{creely2020overview_sparc}. Multi-fidelity training strategies
could combine large volumes of lower-cost data with smaller amounts of
high-fidelity constrained free-boundary solutions.

We also plan to test the resolution-transfer properties of the FNO
explicitly. A single trained operator that evaluates consistently across
grids would interface naturally with multi-fidelity transport, stability,
and control workflows. In addition, physics-aware losses or auxiliary
supervision on derived equilibrium quantities -- finite-difference GS
residual diagnostics, safety-factor proxies, or shape descriptors -- may
help suppress the localized cusp errors that dominate the Hausdorff tail,
and would make the surrogate easier to couple to downstream codes. A
simple ablation of the X-point conditioning would also be a useful
direction for future work, since it would quantify how much of the
geometric fidelity is attributable to explicitly prescribing the
divertor geometry.

This work illustrates a general paradigm for fusion AI: repeated
nonlinear PDE solves over related operating points can be recast as
operator-learning problems, and once the operator has been learned, the
cost of high-fidelity simulation can be amortized without discarding
physically meaningful outputs. Free-boundary equilibrium is a compelling
first target for this paradigm, since both geometry and latency matter
for the use cases that motivate the approach. The ${\sim}640\times$ GPU
and ${\sim}69\times$ CPU speedups reported here -- measured relative to
\textsc{FreeGS} as configured in the present experiments -- combined
with near-deterministic evaluation time (p95/median~$=1.01$), demonstrate
that FNO surrogates can play a practical role in accelerating equilibrium
evaluation in modern fusion workflows, including settings in which
dedicated GPU hardware is unavailable.

\section{Summary and Outlook}
\label{sec:summary}

We have demonstrated millisecond-scale neural-operator surrogate modeling
for constrained double-null free-boundary Grad--Shafranov equilibria. The
FNO learns a constrained forward map from spatial coordinates, scalar
operating parameters, and prescribed X-point geometry to the full
poloidal-flux field on a $65{\times}65$ grid, trained on \textsc{FreeGS}
data for a single fixed machine geometry and prescribed topology with
100\% solver acceptance rate across all 6000 candidate equilibria.

On 500 test equilibria, the best model achieves a mean relative
$L^2$ error of $0.052\%$ and a physical RMSE of
$1.54\times10^{-5}$~Wb. It recovers the separatrix to within 0.072~cm
mean closest-point deviation, both X-points to sub-0.2~cm, and the O-point
to 0.031~cm. Test error follows an empirical $N^{-0.68}$ power law over
$N_{\mathrm{train}} \in \{500, 1000, 2000, 5000\}$, suggesting systematic
improvement with training-set size over the tested regime. The predicted
fields satisfy the external finite-difference GS residual diagnostic at
the same level as the ground-truth \textsc{FreeGS} fields, with a mean
normalized residual of $2.29$, indistinguishable from the
$2.29\pm0.06$ baseline obtained using the same finite-difference scheme.

GPU inference takes $2.77$~ms per equilibrium and CPU inference takes
$25.6$~ms, corresponding to speedups of ${\sim}640\times$ and
${\sim}69\times$ relative to \textsc{FreeGS} as configured here. Both
FNO variants show near-deterministic latency (p95/median~$=1.01$),
compared to $1.13$ for the iterative solver. These results demonstrate
that neural operators can deliver accurate, geometrically faithful, and
finite-difference physics-consistent equilibrium surrogates for
control-oriented workflows, integrated modeling, and large parametric
studies in magnetic-confinement fusion, within a prescribed topology and
machine geometry. Future directions include extension to multiple
topologies and machine geometries, coupled equilibrium--transport
surrogates, resolution-transfer studies, and physics-aware training losses.

\begin{acknowledgments}
The computations in this paper were run on the FASRC Cannon cluster supported by the FAS Division of Science Research Computing Group at Harvard University.
\end{acknowledgments}

\section*{Data and code availability}
The dataset generation script, training code, and evaluation notebook
used in this work will be made available upon reasonable request to the
corresponding author. The \textsc{FreeGS} solver used for data
generation is publicly available at
\texttt{https://github.com/freegs-plasma/freegs}.

\bibliographystyle{aipnum4-2}
\bibliography{references}

\end{document}